\documentclass[aps,preprint,showpacs,preprintnumbers,amsmath,amssymb,floatfix]{revtex4}
\usepackage{graphicx}

\def\beq{\begin{equation}}
\def\eeq{\end{equation}}
\def\beqn{\begin{eqnarray}}
\def\eeqn{\end{eqnarray}}

\begin{document}

\title{Quantum dynamics of attractive versus repulsive bosonic Josephson junctions: 
Bose-Hubbard and full-Hamiltonian results}
\author{Kaspar Sakmann\footnote{E-mail: Kaspar.Sakmann@pci.uni-heidelberg.de}, 
Alexej I. Streltsov\footnote{E-mail: Alexej.Streltsov@pci.uni-heidelberg.de}, 
Ofir E. Alon\footnote{E-mail: Ofir.Alon@pci.uni-heidelberg.de}, 
and Lorenz S. Cederbaum\footnote{E-mail: Lorenz.Cederbaum@pci.uni-heidelberg.de}}
\affiliation{Theoretische Chemie, Physikalisch-Chemisches Institut, Universit\"at Heidelberg,\\
Im Neuenheimer Feld 229, D-69120 Heidelberg, Germany}
\begin{abstract}
The quantum dynamics of one-dimensional 
bosonic Josephson junctions with attractive and repulsive
interparticle interactions 
is studied using the Bose-Hubbard model 
and by numerically-exact computations of the full many-body Hamiltonian.
A symmetry present in the Bose-Hubbard Hamiltonian
dictates an equivalence between
the evolution in time of 
attractive and repulsive Josephson junctions 
with attractive and repulsive interactions of equal magnitude.
The full many-body Hamiltonian does not possess this symmetry
and consequently the 
dynamics of the 
attractive and repulsive junctions 
are different.
\end{abstract}
\pacs{05.60.Gg, 03.65.-w, 03.75.Kk, 03.75.Lm}

\maketitle

Quantum dynamics of interacting Bose-Einstein condensates is an active
and lively research field \cite{pitaevskii_book,leggett_book,pethick_book}.
Here, one of the basic problems studied
is the dynamics of 
tunneling
of interacting Bose-Einstein condensates in double-wells, 
which in this context are referred to as bosonic Josephson junctions.
The dynamics of bosonic Josephson junctions
has drawn much attention
both theoretically and experimentally, 
see, e.g., Refs.~\cite{t1,t2,e1,e2,2mode1,att_rep,M_SU2,clee,2mode2,rgat,meso,M_Polish,2mode3,Anna,BJJ} 
and references therein.

In this work we would like to compare the dynamics of one-dimensional 
bosonic Josephson junctions with attractive and repulsive interparticle interactions.
Explicitly,
we will compare systems
where the {\it magnitude} of attractive and repulsive interactions is alike.
We prepare the interacting bosons in one well,
and then monitor the evolution of the systems in time.
We compute and analyze the respective dynamics both within the two-site Bose-Hubbard model
often employed for this problem as well as within the full Hamiltonian of the systems.
The main result of this work, shown both analytically and numerically,
is that within the Bose-Hubbard model
the dynamics of the attractive and repulsive junctions is equivalent.
In contrast,
the dynamics of attractive and repulsive 
junctions computed from the full many-body Hamiltonian
are different from one another.
As a complementary result we provide here for
the first time in literature numerically-exact quantum dynamics of an
attractive Josephson junction,
thus matching our recent calculations on repulsive Josephson junctions \cite{BJJ}.

We begin with the two-site Bose-Hubbard Hamiltonian:
\beq\label{BHHam}
\hat H(U) = -J \left(\hat b_L^\dag \hat b_R + \hat b_R^\dag \hat b_L\right) + \frac{U}{2} 
\left(\hat b_L^\dag \hat b_L^\dag \hat b_L \hat b_L + \hat b_R^\dag \hat b_R^\dag \hat b_R \hat b_R \right).
\eeq
Here and hereafter all quantities are dimensionless.
We remind the reader that the Bose-Hubbard Hamiltonian (\ref{BHHam}) is
derived from the full many-body Hamiltonian
$
\hat {\mathcal H} =  \int dx \hat{\mathbf \Psi}^\dag(x) h(x) \hat{\mathbf \Psi}(x) +
\frac{\lambda_0}{2} \int dx \hat{\mathbf \Psi}^\dag(x)\hat{\mathbf \Psi}^\dag(x)\hat{\mathbf \Psi}(x) \hat{\mathbf \Psi}(x)
$,
where $h(x)$ is the one-body Hamiltonian and 
$\lambda_0$ the interparticle interaction strength.
Explicitly,
by restricting the field operator to a sum of two terms $\hat{\mathbf \Psi}(x) = \hat b_L \phi_L(x) + \hat b_R \phi_R(x)$,
where $\phi_L(x)$ and $\phi_R(x)$ are the left- and right-localized Wannier functions,
substituting into the many-body Hamiltonian $\hat {\mathcal H}$,
neglecting the off-diagonal interaction terms and eliminating the diagonal one-body terms,
one arrives at the Bose-Hubbard Hamiltonian (\ref{BHHam}).
The one-body Hamiltonian reads $h(x)=\frac{1}{2} \frac{\partial^2}{\partial x^2} + V(x)$,
where $V(x)$ is a symmetric double-well potential.
The left- and right-localized Wannier functions are obtained 
as linear combinations of the ground (gerade) and first-exited (ungerade) states of $h(x)$, i.e.,
$\phi_L(x)=\frac{\phi_g(x)+\phi_u(x)}{\sqrt{2}}$ and 
$\phi_R(x)=\frac{\phi_g(x)-\phi_u(x)}{\sqrt{2}}$, 
and are real-valued functions.
The Bose-Hubbard parameters are given by 
$U = \lambda_0 \int_{-\infty}^{\infty} dx \phi_L^4(x)$ 
and $J = -\int_{-\infty}^{\infty} dx \phi_L(x) h(x) \phi_R(x)$.

There is an interesting symmetry connecting the Bose-Hubbard Hamiltonian (\ref{BHHam})
with repulsive $\hat H(U)$ and attractive $\hat H(-U)$ interactions 
of equal magnitude \cite{sym1,sym2}.
Explicitly, defining the unitary operator (transformation)
\beq\label{RSym}
\hat R=\left\{\hat b_L \to \hat b_L, \hat b_R \to - \hat b_R\right\},
\eeq
it is simple to see that \cite{sym1,sym2}
\beq\label{HSym}
 \hat R \hat H(U) \hat R = - \hat H(-U).
\eeq
What is the impact of the symmetry (\ref{RSym}) and the resulting relation (\ref{HSym})
on the evolution in time of attractive and repulsive bosonic Josephson junctions?

We consider a system of $N$ bosons initially prepared as mentioned above in, say, the left well,
$\left|N,0\right> = \frac{1}{\sqrt{N!}} {\left(\hat b^\dag_L\right)}^N \left|{\mathit vac}\right>$.
Its evolution in time is simply given by $e^{-i \hat H(U) t} \left|N,0\right>$.
Then, the ``survival probability'' of finding the bosons in the left well as a function of time 
is defined as $p_L(t;U) = \frac{1}{N}\int_{-\infty}^{0} dx 
\left<N,0\left| e^{+i \hat H(U) t} \left[\hat{\mathbf{\Psi}}^\dag(x) \hat{\mathbf{\Psi}}(x) \right]
e^{-i \hat H(U) t} \right|N,0\right>$,
where the expression $\left<\ldots\right>$ is the system's density.
Plugging into the ``survival probability'' the expansion of the field operator
and after some algebra the final result reads:
\beqn\label{SProb}
& &  p_L(t;U) = \left(1 - \int_{-\infty}^0 dx \phi_L^2(x) \right) -
\left(1 - 2\int_{-\infty}^0 dx \phi_L^2(x) \right) \cdot \frac{1}{N} \nonumber \\
& & \qquad \times
 \left\{ \left<N,0\left| \cos [\hat H(U) t]\hat b_L^\dag \hat b_L \cos [\hat H(U) t] +
\sin [\hat H(U) t]\hat b_L^\dag \hat b_L \sin [\hat H(U) t]
 \right|N,0\right> \right\}. \
\eeqn
In obtaining the r.h.s. of (\ref{SProb})
we made use of the facts that 
the expectation value of $\hat b_L^\dag \hat b_L$ (hermitian operator)
is real,
and that $\int_{-\infty}^0 dx \phi_L(x) \phi_R(x) = 0$.

Employing the $\hat R$ symmetry (\ref{RSym}) to 
the $p_L(t;U)$ matrix element (\ref{SProb}) and making use of relation (\ref{HSym}),
one immediately finds that
\beq\label{SSur}
p_L(t;-U) = p_L(t;U),
\eeq 
which concludes our first proof.
In other words,
the ``survival probability'' of bosons is identical
for attractive and repulsive interactions (of equal magnitude)
within the Bose-Hubbard model.
We emphasize that the result (\ref{SSur})
holds at all times $t$.

Next, we discuss the impact of the symmetry (\ref{RSym}) 
on the eigenvalues of the reduced one-body density matrix
within the two-site Bose-Hubbard model.
The eigenvalues of the reduced one-body density matrix of a Bose system
determine the extent to which the system is condensed or fragmented \cite{MCHB,ueda}.
For the two-site Bose-Hubbard problem the reduced one-body density matrix 
can be written as a two-by-two matrix:
\beq\label{Red_U}
 \rho^{(1)}(t;U) = \begin{pmatrix} \rho_{LL}(t;U) & \rho_{LR}(t;U) \cr \rho_{LR}^\ast(t;U) & \rho_{RR}(t;U) \cr \end{pmatrix},
\eeq
where $\rho_{LL}(t;U) = \left<N,0\left| e^{+i \hat H(U) t} \hat b_L^\dag \hat b_L  e^{-i \hat H(U) t} \right|N,0\right>$,
and $\rho_{LR}(t;U)$ and $\rho_{RR}(t;U)$ are given analogously.
Plugging the symmetry (\ref{RSym}) 
into each of the matrix elements of $\rho^{(1)}(t;U)$
and making use of relation (\ref{HSym}),
one can straightforwardly express the 
reduced one-body density matrix for attractive interaction as follows:
\beq\label{Red_mU}
 \rho^{(1)}(t;-U) = \begin{pmatrix} \rho_{LL}(t;U) & -\rho_{LR}^\ast(t;U) \cr 
 -\rho_{LR}(t;U) & \rho_{RR}(t;U) \cr \end{pmatrix}.
\eeq
Obviously, 
the matrices (\ref{Red_U}) and (\ref{Red_mU})
have the same characteristic equation, 
and hence the same eigenvalues.
We have thus shown that, 
within the Bose-Hubbard model,
the eigenvalues of the reduced one-body density matrix
do not depend on the sign of interparticle interaction,
which constitutes our second proof.
Again, this result holds 
for any time $t$.
In particular,
the Bose-Hubbard model attributes identical condensation and fragmentation levels
to attractive and repulsive systems.

\begin{figure}
\includegraphics[width=8.6cm, angle=0]{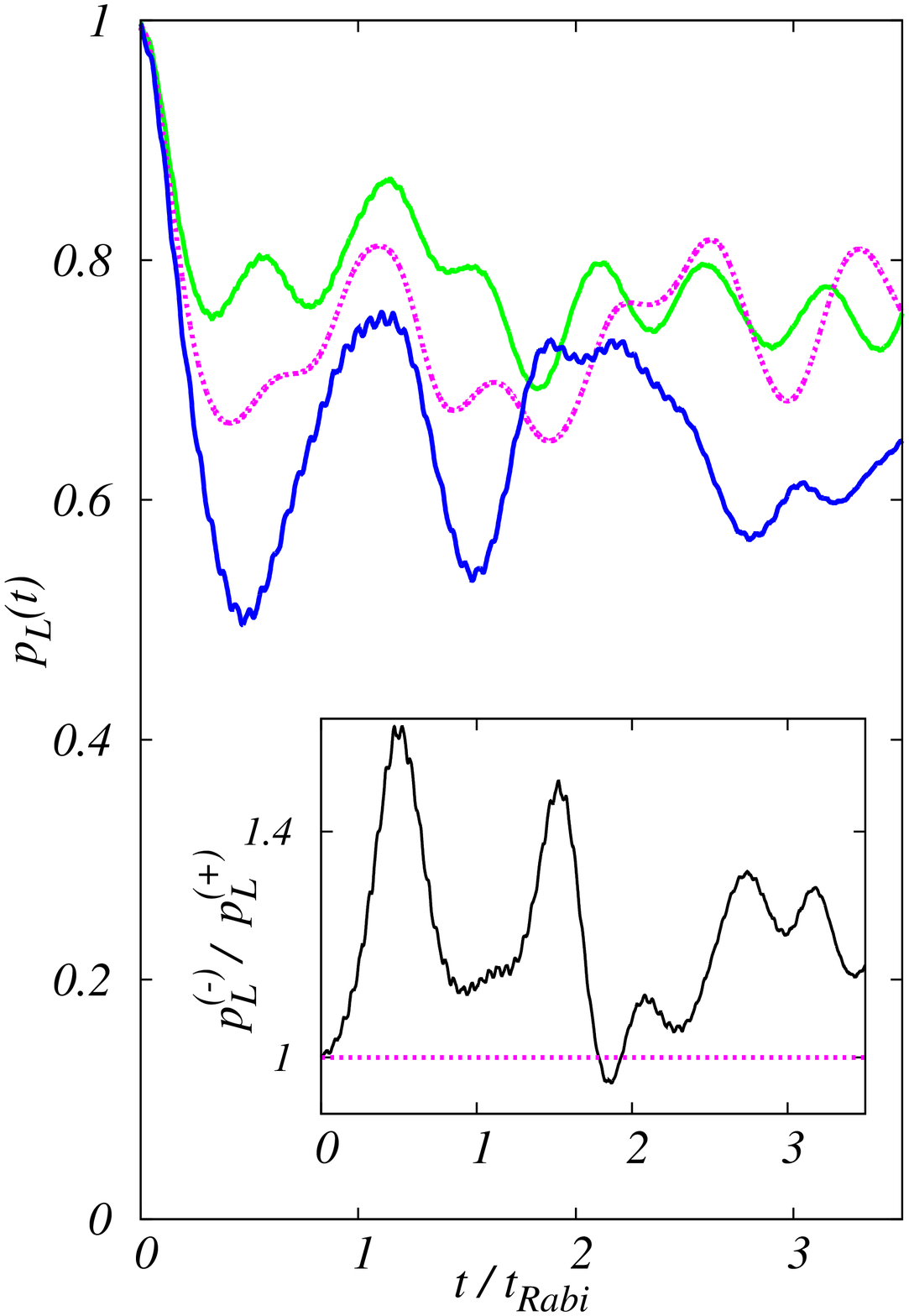}
\caption {(Color online) Bose-Hubbard versus full-Hamiltonian, numerically-exact dynamics
of attractive and repulsive Josephson junctions.
Shown is the ``survival probability'' as a function of time, $p_L(t)$,
computed with the full many-body Hamiltonian for attractive
[solid gray (green) line] and repulsive [solid black (blue) line] interaction.
The Bose-Hubbard result [dashed (magenta) line]
is for {\it both} attractive and repulsive interactions.
The parameters used are 
$N=20$, $\frac{|U|}{J}=0.226$, $|\lambda_0|=0.0129$, and $t_{\mathit Rabi}=140.66$.
The inset shows the ratio $p^{(-)}_L/p^{(+)}_L$
of attractive to repulsive ``survival probabilities'' as a function of time.
The black--solid line is the full-Hamiltonian results which exhibit a complex
dynamics, 
whereas the dashed--magenta 
line is the Bose-Hubbard result,
showing no dynamics at all.
All quantities are dimensionless.
}
\label{fig1}
\end{figure}

\begin{figure}[ht]
\includegraphics[width=8.6cm, angle=0]{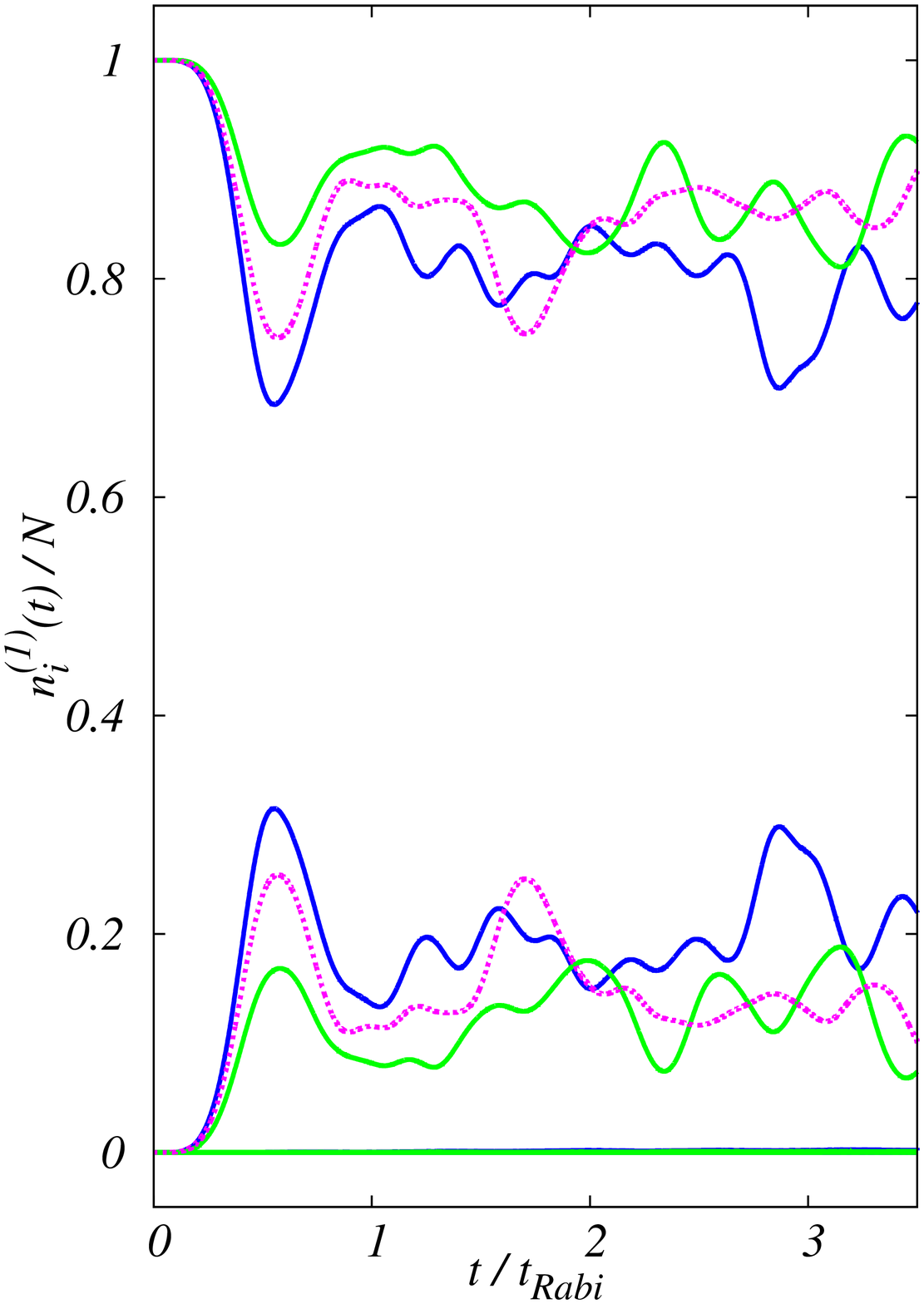}
\caption [kdv]{(Color online) 
Bose-Hubbard versus full-Hamiltonian, numerically-exact dynamics
of attractive and repulsive Josephson junctions.
Shown are the occupation numbers of the reduced one-body density matrix 
as a function of time, $n^{(1)}_i(t)$,
computed with the full many-body Hamiltonian for attractive
[solid gray (green) line] and repulsive [solid black (blue) line] interaction.
The Bose-Hubbard result [dashed (magenta) line]
is for {\it both} attractive and repulsive interactions.
The two-site Bose-Hubbard dynamics has two occupation numbers only.
The full-Hamiltonian dynamics has many occupation numbers.
It is seen that the 
occupation numbers $n^{(1)}_{i>2}(t)$
are essentially zero.
The parameters used are the same as in Fig.~\ref{fig1}.
All quantities are dimensionless.
}
\label{fig2}
\end{figure}

To illustrate the above findings
we plot in Figs.~\ref{fig1} and \ref{fig2} 
the ``survival probability'' and occupation numbers, respectively, 
as a function of time for a repulsive and attractive
Bose-Hubbard dynamics with the ratio of 
parameters $\frac{|U|}{J}=0.226$.
These parameters draw from a double-well potential $V(x)$ formed 
by connecting two harmonic potentials $V_{\pm}(x)=\frac{1}{2}(x\pm 2)^2$
with a cubic spline in the region $|x|\le 0.5$,
and for $|\lambda_0|=0.0129$. 
The number of bosons is $N=20$. 
The Rabi oscillation period is $t_{\mathit Rabi}=\pi/J=140.66$.
As expected, 
based on Eq.~(\ref{SSur}) 
and Eqs.~(\ref{Red_U},\ref{Red_mU}),
the Bose-Hubbard dynamics for attractive and repulsive junctions
are identical.

We now move on to the dynamics computed with the full many-body Hamiltonian $\hat {\mathcal H}$.
Recently, we have reported on the numerically-exact solution of a one-dimensional
repulsive bosonic Josephson junction,
which has allowed us to unveil novel phenomena associated with 
the quick loss of the junction's coherence \cite{BJJ}.
We use here the same method 
to compute the time evolution of the system with the full many-body Hamiltonian $\hat {\mathcal H}$.
We solve the time-dependent many-boson Schr\"odinger equation by
using the multiconfigurational time-dependent 
Hartree for bosons (MCTDHB) method \cite{MCTDHB} (also see \cite{fragmenton,Joerg}),
in which a novel mapping of the many-boson configuration space in 
combination with a parallel implementation of MCTDHB are exploited \cite{parll}.
This allows us to report, among others,
on the first numerically-exact results in literature of 
a bosonic Josephson junction for attractive interaction,
thus matching our recent calculations on repulsive bosonic Josephson junctions \cite{BJJ}.
The results of the computations with the full many-body 
Hamiltonian are collected in Figs.~\ref{fig1} and \ref{fig2}.
It is clearly seen that 
the dynamics of the attractive and repulsive junctions 
are distinct from each other,
and that each is different from the Bose-Hubbard dynamics.

Let us analyze these findings.
We first note, in the context of the above analytical results on the Bose-Hubbard dynamics,
that the full Hamiltonian $\hat {\mathcal H}$
does not possesses the symmetry (\ref{RSym})
connecting 
the dynamics of attractive and repulsive systems.
This is because $\hat {\mathcal H}$ contains off-diagonal interaction terms as well
as all other terms neglected in the Bose-Hubbard Hamiltonian (\ref{BHHam}).
From this ``mathematical'' discussion alone,
we do not expect the dynamics of attractive and repulsive junctions
to be equivalent as found above for the Bose-Hubbard dynamics.
What do we expect on physical grounds?
Intuitively, we know that 
attractive bosons like to be together,
whereas repulsive bosons lean 
to separate from one another.
These tendencies are exactly what we see in Fig.~\ref{fig1}.
The full-Hamiltonian's ``survival probability'' is
larger (smaller) 
for attractive (repulsive) interaction
than the Bose-Hubbard ``survival probability'',
at least up to $t/t_{\mathit Rabi}=1.5$.
In other words,
the Bose-Hubbard ``survival probability'' underestimates the ``survival probability'' for attractive 
and overestimates it for repulsive interaction 
for short and intermediate times.
For longer times, as seen in Fig.~\ref{fig1},
the dynamics becomes more complex and anticipating 
the differences between the exact and the 
Bose-Hubbard dynamics
cannot rest on the 
above-mentioned physical intuition alone.
Finally, 
Fig.~\ref{fig2} presents a complementary picture of the dynamics of occupation numbers.
It has been shown in \cite{BJJ}
that the Bose-Hubbard dynamics underestimates
fragmentation and overestimates coherence
of repulsive bosonic Josephson junctions.
We may analogously anticipate that the reverse happens with attractive interactions,
which indeed is the physical 
picture unveiled in Fig.~\ref{fig2}.

Let us conclude.
We have shown, both analytically and numerically,
that a symmetry present in the two-site Bose-Hubbard Hamiltonian
dictates an equivalence between
the evolution in time of attractive and repulsive bosonic Josephson junctions.
The full many-body Hamiltonian does not possess this symmetry
and consequently the dynamics of the 
attractive and repulsive junctions are 
distinct.
The Bose-Hubbard dynamics underestimates the ``survival probability'' and overestimates
fragmentation of attractive one-dimensional bosonic Josephson junctions
and the reverse is true for repulsive junctions.
Note that the parameters used here are within 
the range of expected validity of the Bose-Hubbard model for Josephson junctions \cite{t1}.
The clear deviations from the
numerically-exact results show that criteria for the validity
of the Bose-Hubbard model which have been derived for static junctions cannot be transferred
for dynamically evolving junctions (also see \cite{BJJ}).
The present investigation of attractive versus repulsive junctions sheds
additional light on the
restrictions of the Bose-Hubbard model to describe dynamics.

As an outlook,
we mention that an analogous symmetry to (\ref{RSym})
can be found for the multi-site Bose-Hubbard model.
Consider the multi-site one-dimensional Bose-Hubbard model:
$\hat H_{\mathit BH}(U) = -J \sum_j \hat b_j^\dag \hat b_{j+1} + 
\frac{U}{2}\sum_j \hat b_j^\dag \hat b_j^\dag \hat b_j \hat b_j$.
Then, $\hat R_{\mathit BH} \hat H_{\mathit BH}(U) \hat R_{\mathit BH} = - \hat H_{\mathit BH}(-U)$
where $\hat R_{\mathit BH}=\left\{\hat b_{2j} \to \hat b_{2j}, \hat b_{2j-1} \to - \hat b_{2j-1}\right\}$.
The extension to the Bose-Hubbard model
of orthorhombic lattices in higher dimensions 
is straightforward.
It would be interesting to search for the consequences
of this symmetry 
in 
the dynamics of attractive and repulsive 
bosons in a lattice. 

Financial support by the DFG and 
computing time at the bwGRiD and HELICS II are acknowledged.

\begin{thebibliography}{99}


\bibitem{pitaevskii_book} L. Pitaevskii and S. Stringari,
                          {\it Bose-Einstein Condensation}
                          (Oxford University Press, Oxford, 2003).

\bibitem{leggett_book} A. J. Leggett,
                       {\it Quantum Liquids: Bose condensation and Cooper pairing in condensed matter systems}
                       (Oxford University Press, Oxford, 2006).

\bibitem{pethick_book}  C. J. Pethick and H. Smith,
                        {\it Bose-Einstein Condensation in Dilute Gases}, 2nd ed.
                        (Cambridge University Press, Cambridge, 2008).

\bibitem{t1} G. J. Milburn, J. Corney, E. M. Wright, and D. F. Walls,
             Phys. Rev. A {\bf 55}, 4318 (1997).

\bibitem{t2} A. Smerzi, S. Fantoni, S. Giovanazzi, and S. R. Shenoy, 
             Phys. Rev. Lett. {\bf 79}, 4950 (1997).

\bibitem{e1} M. Albiez, R. Gati, J. F\"olling, S. Hunsmann, M. Cristiani, and M. K. Oberthaler,
             Phys. Rev. Lett. {\bf 95}, 010402 (2005).  

\bibitem{e2} S. Levy, E. Lahoud, I. Shomroni, and J. Steinhauer,
             Nature (London), {\bf 449}, 579 (2007).

\bibitem{2mode1} S. Raghavan, A. Smerzi, S. Fantoni, and S. R. Shenoy,
                 Phys. Rev. A {\bf 59}, 620 (1999).

\bibitem{att_rep} E. A. Ostrovskaya, Y. S. Kivshar, M. Lisak, B. Hall, F. Cattani, and D. Anderson,
                  Phys. Rev. A {\bf 61}, 031601(R) (2000). 

\bibitem{M_SU2} Y. Zhou, H. Zhai, R. L\"u, Z. Xu, and L. Chang,
                Phys. Rev. A {\bf 67}, 043606 (2003). 

\bibitem{clee} C. Lee, 
               Phys. Rev. Lett. {\bf 97}, 150402 (2006).

\bibitem{2mode2} D. Ananikian and T. Bergeman,
                 Phys. Rev. A {\bf 73}, 013604 (2006).

\bibitem{rgat} R. Gati and M. K. Oberthaler, 
               J. Phys. B {\bf 40}, R61 (2007).

\bibitem{meso} G. Ferrini, A. Minguzzi, and F. W. J. Hekking,
               Phys. Rev. A {\bf 78}, 023606 (2008).

\bibitem{M_Polish} V. S. Shchesnovich and M. Trippenbach,
                   Phys. Rev. A {\bf 78}, 023611 (2008).

\bibitem{2mode3} X. Y. Jia, W. D. Li, and J. Q. Liang,
                 Phys. Rev. A {\bf 78}, 023613 (2008). 

\bibitem{Anna} M. Trujillo-Martinez, A. Posazhennikova, and J. Kroha,
               Phys. Rev. Lett. {\bf 103}, 105302 (2009).

\bibitem{BJJ} K. Sakmann, A. I. Streltsov, O. E. Alon, and L. S. Cederbaum,
              Phys. Rev. Lett. {\bf 103}, 220601 (2009).

\bibitem{sym1} J. Links and K. E. Hibberd,
               SIGMA {\bf 2} (2006), Paper 095.

\bibitem{sym2} J. Links and S.-Y. Zhao,
               J. Stat. Mech. (2009) P03013.

\bibitem{MCHB} A. I. Streltsov, O. E. Alon, and L. S. Cederbaum, 
               Phys. Rev. A {\bf 73}, 063626 (2006). 

\bibitem{ueda} E. J. Mueller, T.-L. Ho, M. Ueda, and G. Baym,
               Phys. Rev. A {\bf 74}, 033612 (2006).


\bibitem{MCTDHB} A. I. Streltsov, O. E. Alon, and L. S. Cederbaum, 
                 Phys. Rev. Lett. {\bf 99}, 030402 (2007);
                 O. E. Alon, A. I. Streltsov, and L. S. Cederbaum, 
                 Phys. Rev. A {\bf 77}, 033613 (2008).

\bibitem{fragmenton} A. I. Streltsov, O. E. Alon, and L. S. Cederbaum,
                     Phys. Rev. Lett {\bf 100}, 130401 (2008).

\bibitem{Joerg} J. Grond, J. Schmiedmayer, and U. Hohenester,
                Phys. Rev. A {\bf 79}, 021603(R) (2009);
                J. Grond, G. von Winckel, J. Schmiedmayer, and U. Hohenester,
                arXiv:0908.1634v1.

\bibitem{parll} A. I. Streltsov, O. E. Alon, and L. S. Cederbaum, arXiv:0910.2577v1;
                A. I. Streltsov, K. Sakmann, O. E. Alon, and L. S. Cederbaum, arXiv:0910.5916v1.

\end{thebibliography}
\end{document}